\documentclass[a4paper, 9 pt, conference]{ieeeconf} 

\IEEEoverridecommandlockouts                              
\usepackage{cite}
\usepackage{graphicx} 
\usepackage{cuted}
\usepackage[switch]{lineno}
\usepackage{amsmath} 
\usepackage{amssymb}
\usepackage{mathpazo}
\usepackage[mathpazo]{flexisym}
\usepackage{breqn}

\newcommand{\be}{\begin{equation}}
\newcommand{\ee}{\end{equation}}
\newcommand{\beqn}{\begin{eqnarray}}
\newcommand{\eeqn}{\end{eqnarray}}
\newcommand{\beginsupplement}{%
     \setcounter{page}{1}
         \renewcommand{\thepage}{S\arabic{page}}%
        \setcounter{section}{0}
         \renewcommand{\thesection}{S}%
        \setcounter{table}{0}
        \renewcommand{\thetable}{S\arabic{table}}%
        \setcounter{figure}{0}
        \renewcommand{\thefigure}{S\arabic{figure}}%
          \setcounter{equation}{0}
         \renewcommand{\theequation}{S\arabic{equation}}%
     }

\title{\LARGE \bf
Cross-scale cooperation enables sustainable use of a common-pool resource}

\author{Andrew K. Ringsmuth$^{1,*}$, Steven J. Lade$^{1,2}$ and Maja Schl\"uter$^{1}$%
\thanks{$^{1}$Stockholm Resilience Centre, Stockholm University, Kr\"aftriket 2B, 10691, Stockholm, Sweden}
\thanks{$^*$a.k.ringsmuth@gmail.com}%
\thanks{$^{2}$Fenner School of Environment \& Society, The Australian National University, Building 141, Linnaeus Way, Canberra, ACT, 2601, Australia}%
}

\begin{document}

\maketitle
\thispagestyle{plain}
\pagestyle{plain}

\begin{abstract}
In social-ecological systems (SESs), social and biophysical dynamics interact within and between structural levels separated by spatial and temporal scales. Cross-scale interactions (CSIs) are interdependences between processes at different scales, generating behaviour unpredictable at single scales. Understanding CSIs is important for improving SES governance but they remain understudied. Theoretical models are needed, which capture essential features while being simple enough to yield insights into mechanisms. In a stylised model, we study CSIs in a two-level system of weakly interacting communities harvesting a common-pool resource. Community members adaptively conform to, or defect from, a norm of socially optimal harvesting, enforced through social sanctioning both within and between communities. Each subsystem's dynamics depend sensitively on the other despite interactions being much weaker between subsystems than within them. When interaction is purely biophysical, stably high cooperation in one community can cause cooperation in the other to collapse. However, even weak social interaction can prevent collapse of cooperation and instead cause collapse of defection. We identify conditions under which subsystem-level cooperation produces desirable system-level outcomes. Our findings expand evidence that collaboration is important for sustainably managing shared resources, showing its importance even when resource sharing and social relationships are weak.\\

\end{abstract}


\section{INTRODUCTION}

In social-ecological systems (SESs), social and biophysical dynamics interact within and between levels of organisation separated by spatial and temporal scales. The Anthropocene is characterised by global changes emerging from local changes in human-environment interactions. Global changes in turn cause local impacts and most attempts to manage them also must be deployed locally \cite{verburg_2016}. The multiscale nature of SES dynamics and governance makes theoretical characterisation of cross-scale dynamics important, to help understand how changes implemented at a given level will propagate across scales \cite{mcguire2018,chaffin2016,soranno2014,folke2010,ostrom2009,cash2006,folke2005,peters2004,gibson2000}. 

 According to hierarchy theory, when levels are separated by scale, a complex system is `quasiseparable' into semi-independent levels; when modelling one level, variables at other levels may be treated as constant boundary conditions \cite{gibson2000,giampietro1994}. In such systems, causes of nonlinear behaviour at one level, such as threshold behaviour, must lie at the same level. However, SESs are not separable in this way. If a variable at some level approaches a critical threshold, even a small perturbation from an interaction with a variable at another level may drive it over the threshold, causing nonlinear behaviour such as a regime shift \cite{walker2012,peters2007,gibson2000}. Such sensitive dependence between processes at different scales, which generates behaviour unpredictable from behaviour at single scales, is known as a cross-scale interaction (CSI) \cite{peters2007}. Many researchers have qualitatively described CSIs in SESs, conceptually \cite{isbell2017,cumming2015,maciejewski2015,scholes2013,walker2012,brondizio2009,cash2006,gibson2000} and in specific case studies (recently, \cite{mcguire2018,niiranen2018,devos2017,cox2014,heikkila2011}). CSIs have been statistically quantified in different systems, from data at multiple scales. Brondizio \emph{et al} \cite{brondizio2012} analysed Amazon deforestation data at different length scales, showing that understanding deforestation trajectories requires differentiating causes at different scales. Soranno and collaborators \cite{soranno2014} quantified CSIs using multiscale data and Bayesian hierarchical statistical models. Applying this method to a study of lake water quality across different regions of North America, they identified a scale mismatch between quality variation and water management.

To date, however, there have been few attempts to theoretically model CSIs in SESs in terms of underlying mechanisms. Peters \emph{et al} \cite{peters2004} proposed a mathematical framework for nonlinear dynamics in catastrophic events in terms of pattern-process relationships, including feedbacks, across scales. They applied this framework to characterise events such as the spread of wildfires, infectious diseases and insect outbreaks, finding that strategies for mitigating risks of such events must account for CSIs and will often be counterintuitive. Recently, Lansing and coworkers reported \cite{lansing2017} an agent-based model of water resource and pest control management in Bali's ancient rice terraces, to help explain shifts from individualistic field-level practices to cooperative regional-level practices. Modelling agent interactions and management decisions using evolutionary game theory on a spatially embedded lattice model showed that long-established spatial patterns observable in the rice terraces can be created by feedback between farmers' decisions and the ecology of the paddies, triggering a transition from individualistic to cooperative practices. In the theoretical middle ground between generic frameworks and detailed case models lies potential for stylised, minimal models. These aim to capture essential system features while being simple enough to yield insights into mechanisms. This is helpful for guiding thought experiments relating CSIs to underlying mechanisms, and for interpreting empirical case study results.  

Here, we develop a stylised model of resource management by social pressure in a multilevel SES, accounting for CSIs. Focusing on common-pool resource management, we extend prior work by Tavoni, Schl\"uter and Levin (TSL) \cite{tavoni2012}, who used evolutionary game theory to model a community of agents harvesting a renewable common-pool resource as a basis for economic production. Agents adaptively chose whether to conform to, or defect from, a norm of socially optimal harvesting, enforced through social sanctioning. Schl\"uter \emph{et al} \cite{schluter2016} extended the TSL model, using agent-based methods to test the robustness of norm-driven cooperation to environmental variations. One was the introduction of a neighbouring community and resource pool, with diffusive resource transfer between the pools. Results showed that stably high cooperation in one community could cause cooperation in the other to collapse because added resource availability due to the pools' coupling increased the payoff for norm defectors. However, the communities did not adjust their norms for added resource availability, and no social interactions between communities were considered. 

We extend this analysis of interacting SESs with two types of social interaction and characterise how local-level behaviours interact to produce outcomes at the system level. First, we consider how communication between communities while setting harvesting norms can adjust the norms for added resource availability due to biophysical coupling. Second, we add ongoing social coupling such that norm-cooperative agents can sanction defectors in both communities. We assume separation of scales between the strengths of interactions within and between subsystems. This enables us to quantify how nonlinear changes in subsystem behaviour (appearance or disappearance of dynamical attractors) arise from biophysical and social CSIs between the two subsystems. We identify conditions under which subsystem-level cooperation gives rise to desirable system-level outcomes, and conditions which prevent this.

The exact system model is analytically insoluble and we use perturbation theory \cite{shivamoggi2003} to derive an approximate model in the weak coupling (multiscale) regime. This model closely approximates numerical solutions of the exact model. Solving our model with a combination of analytical and numerical methods allows more efficient characterisation of the parameter space than purely numerical methods, and the derived equations shed light on mechanisms. This dual approach is also a basis for future development using more advanced multiscale analytical methods from dynamical systems theory and statistical physics, which have elucidated a wide range of complex systems \cite{attard2012,mccomb2007,kevorkian2012,pesenson2013,longo2013}. We envisage that, in future, a combination of mathematical and agent-based stylised multiscale models may enable an effective theory of CSIs in SESs, complementing statistical and qualitative analyses from multiscale empirical studies. This may help to guide empirical researchers in quantifying CSIs and mechanisms underlying them, revealing new leverage points for governance of real systems.

\section{MODEL}
\begin{figure}[ht]
\includegraphics[angle=0,width=\columnwidth]{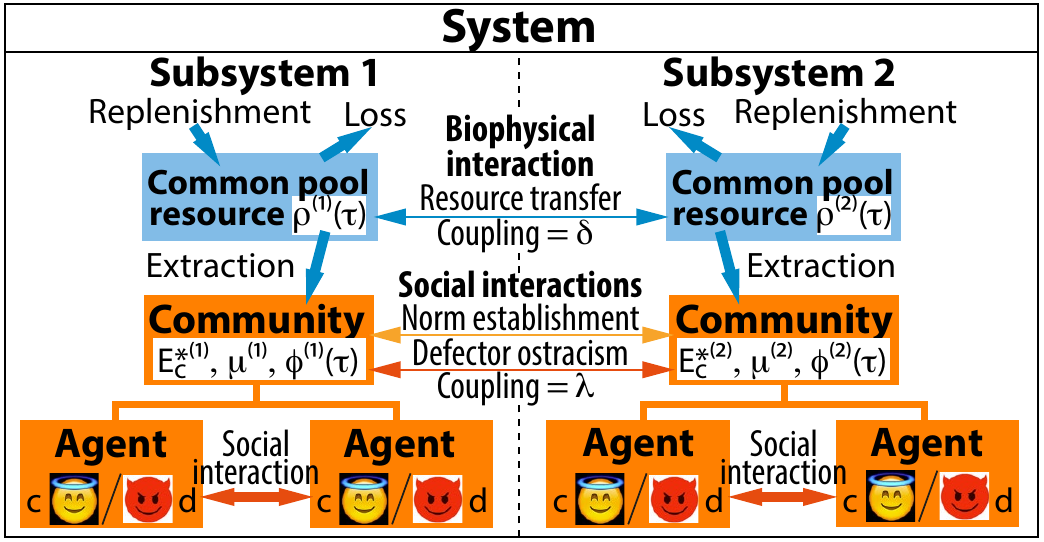}\centering
\caption{\textbf{Model system with social and biophysical interactions between symmetrical subsystems.} In each, a community of agents extracts a renewable common pool resource and each adaptively chooses between cooperative (low) and defective (high) extraction effort levels, constrained by a social norm of non-excessive harvesting (quantified by ${E^*_c}^{(i)}$). Cooperators enforce the norm by ostracising defectors. Biophysical coupling (strength $\delta$) permits resource transfer between subsystems. Communities may choose to share resource information while setting norms; we assume that this communication is either perfect or nonexistent. Ongoing social coupling (strength $\lambda$) enables cooperators to ostracise defectors in both communities.}
\label{fig:syspair}
\end{figure}

We consider the two-level model schematised in figure \ref{fig:syspair}, comprising two interacting social-ecological subsystems, each of which, in isolation, is described by the TSL model \cite{tavoni2012}. In each, a community of agents harvests a common-pool resource. Each agent adaptively conforms to, or defects from, a social norm of socially optimal harvesting, choosing a strategy based on resource availability and the utilities of strategies chosen by other agents. The biophysical variables of interest are the resource stocks, and social variables the norm-cooperative fractions of the subsystem populations. The two are directly coupled only within each subsystem (agents harvest only locally), but the biophysical and social variables are coupled between subsystems. Social interaction between subsystems comprises two steps: first, communities may choose to share information about their resources while establishing harvesting norms; second, cooperators ostracise defectors to enforce the norms. 

\subsection{Biophysical dynamics}\label{subsec:resdyn}
  We model subsystem resource dynamics as the sum of local processes and subsystem interactions. Following TSL \cite{schluter2016,tavoni2012}, the resource stock of, for example, subsystem 1 ($R^{(1)}(t)$) is replenished by its environment at constant rate ($c^{(1)}$) and lost at a rate that depends on the square of the resource occupation (ratio of the stock to its maximum value, $R_{m}^{(1)}$, determined by limits such as the capacity of a reservoir). Resource is also harvested by the subsystem community of $n^{(1)}$ agents, collectively exerting effort $E^{(1)}(t)$. Resource is transferred diffusively between subsystems such that the rate depends linearly on the difference between their resource occupations. Under the combined action of these processes, subsystem 1's resource stock changes according to the first-order, nonlinear ordinary differential equation, 
  \beqn
  \dot{R}^{(1)}(t)&=&c^{(1)}-d^{(1)}\left(\frac{R^{(1)}(t)}{R_{m}^{(1)}}\right)^2-q E^{(1)}(t) R^{(1)}(t)\nonumber\\
  &~&~~~~~~~~~~~~~~~+r\left(\frac{R^{(2)}(t)}{R_{m}^{(2)}}-\frac{R^{(1)}(t)}{R_{m}^{(1)}}\right).\label{eqn:dR}
  \eeqn
Here, $d^{(1)}$ is a coefficient particular to a given system and $q$ (the `technology factor') is a constant. In real systems, resource transfer coupling, $r$ may depend on flow direction due to, for example, constraints of geography or institutional regulations. However, for simplicity we assume that $r$ is independent of flow direction and that both subsystems have equal resource capacities, and replenishment and dissipation coefficients ($R_{m}^{(i)}\equiv R_{m}$, $c^{(i)}\equiv c$, $d^{(i)}\equiv d$). So that the effects of biophysical and social subsystem interactions may be compared fairly, we define dimensionless variables, $\rho^{(i)}=R^{(i)}/R_m$ and $\tau=(d/R_m)t$, and rewrite (\ref{eqn:dR}) as

\beqn
\dot{\rho}^{(1)}(\tau)=\frac{c}{d}-\rho^{(1)}(\tau)^2-\frac{q R_m}{d}E^{(1)}(\tau) \rho^{(1)}(\tau)~~~~\nonumber\\~~~~~~~~~~~~~~~~~~~~~~~+\frac{r}{d}(\rho^{(2)}(\tau)-\rho^{(1)}(\tau)).\label{eqn:drho}
\eeqn

In a given system, it may be possible to quantify how coupling strength depends on space and time. For example, if communities harvest water from connected bodies, the mechanics of water transfer between them will depend on geographical parameters such as distance. However, for generality, we impose separation of scales between subsystem- and system-level interaction strengths phenomenologically, by invoking a weak coupling parameter, $\delta\equiv \frac{r}{d}, ~0\leq\delta\ll 1$. This allows us to treat system-level resource transfer as a small perturbation to subsystem-level dynamics and, therefore, using regular perturbation theory \cite{shivamoggi2003}, to assume that (\ref{eqn:drho}) has solutions of the form 

\be
\rho^{(1)}(\tau)=\rho_0^{(1)}(\tau)+\delta \rho_1^{(1)}(\tau)+\delta^2 \rho_2^{(1)}(\tau)+....\label{eqn:solnp}
\ee

Substituting (\ref{eqn:solnp}) into (\ref{eqn:drho}) and truncating at first order in $\delta$ yields differential equations for the isolated subsystem resource dynamics and a perturbation due to system-level resource transfer, respectively:
\begin{subequations}
\beqn
\dot{\rho_0}^{(1)}(\tau)&=&\frac{c}{d}-\rho_0^{(1)}(\tau)^2-\frac{q R_m}{d}E^{(1)}(\tau) \rho_0^{(1)}(\tau),\label{eqn:drho0}\\
\dot{\rho_1}^{(1)}(\tau)&=&\rho_0^{(2)}(\tau)-\rho_0^{(1)}(\tau)(1+2 \rho_1^{(1)}(\tau))\nonumber\\&{}&~~~~~~~~~~~~~~-\frac{q R_m}{d}E^{(1)}(\tau) \rho_1^{(1)}(\tau).\label{eqn:drho1}
\eeqn
\end{subequations}
In the supporting information (SI), we show that this description very closely approximates the complete description (\ref{eqn:drho}) (see fig. \ref{fig:transient}). Perturbation theory makes the system mathematically tractable and also links it to core concepts of social-ecological resilience \cite{folke2010}, which may be defined as the capacity of a SES to absorb perturbations while maintaining structure and function \cite{chaffin2016, folke2010}. Our analysis directly assesses subsystem resilience to cross-scale perturbations.

\subsection{Social dynamics}
\begin{figure}[ht]
\includegraphics[angle=0,width=0.8\columnwidth]{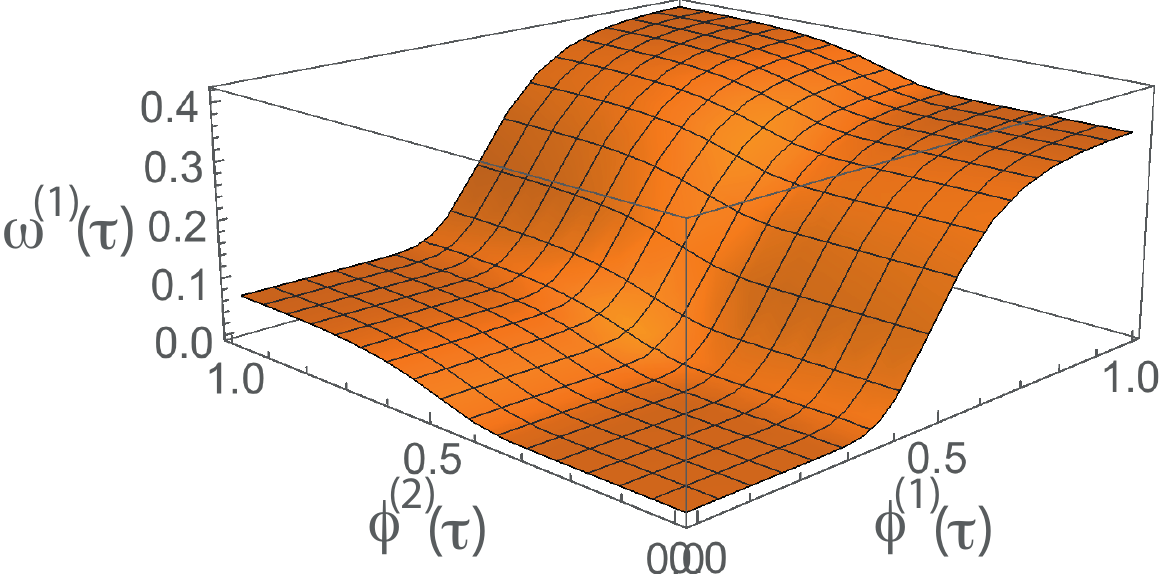}\centering
\caption{\textbf{Ostracism experienced by defectors in subsystem 1 due to the cooperative fractions of the subsystem 1 and 2 populations.} Parameters: $n=50$, $\lambda=0.2$, $s=0.34/n=6.8\times10^{-3}$, $T=-150$, $g=-10$. The $s$ value is derived from the value assumed in \cite{tavoni2012} for the maximum sanctioning applicable by the whole community, $h=0.34$. }
\label{fig:ost}
\end{figure}

Total community extractive effort is the sum of all cooperators' and defectors' efforts. For subsystem 1, 
\beqn
E^{(1)}(\tau)=n^{(1)}{e_c}^{(1)}\left(\mu^{(1)}+(1-\mu^{(1)})\phi^{(1)}(\tau)\right).\label{eqn:Ef}
\eeqn
Here, $0\leq \phi^{(1)}(\tau)\leq1$ is the population fraction made up of norm cooperators, who each extract at socially optimal (lower) effort level, ${e_c}^{(1)}$, while the remaining defectors exert (higher) ${e_d}^{(1)}$. We define $\mu^{(1)}={e_d}^{(1)}/{e_c}^{(1)}$. Community productivity is assumed to take Cobb-Douglas form in effort and resource \cite{tavoni2012}:
\beqn
f^{(1)}(\tau)&=&\gamma [E^{(1)}(t)]^\alpha [R^{(1)}(t)]^\beta\nonumber\\
&=&\gamma R_m^\beta [E^{(1)}(\tau)]^\alpha [\rho^{(1)}(\tau)]^\beta, \label{eqn:prod}
\eeqn
where $\gamma$ is the total factor productivity and $\alpha$, $\beta$ are constants. Payoffs for cooperators ($c$) and defectors ($d$) are \cite{tavoni2012}
\beqn
{\pi_x}^{(1)}(\tau)={e_x}^{(1)}\left(\frac{f^{(1)}(\tau)}{E^{(1)}(\tau)}-w\right), ~~~x=c,d,\label{eqn:pay}
\eeqn
where $w$ is the cost of production. Each defector experiences social sanctioning (`ostracism') from the cooperative fractions of both communities, weighted by subsystem populations and social couplings, $\lambda^{(ij)}$: 

\beqn
\omega^{(1)}(\tau)&=&s\left(n^{(1)}e^{T e^{g \phi^{(1)}(\tau)}}+\lambda^{(21)} n^{(2)}e^{T e^{g \phi^{(2)}(\tau)}}\right),~~~\label{eqn:ost}
\eeqn
where $s$ is the maximal sanctioning that an individual can apply. We assume symmetrical community populations and social couplings ($n^{(i)}\equiv n$; $\lambda^{(ij)}\equiv\lambda$), but note that in real systems they may differ for many reasons. Mechanisms underlying social coupling are particular to a given system. For generality, we again impose separation of scales by assuming that couplings are much weaker between individuals in separate communities than in the same community ($0\leq\lambda\ll 1$). Figure \ref{fig:ost} plots the ostracism experienced by a defector in subsystem 1 due to cooperators in both subsystems. The dependences on $\phi^{(1)}$ and $\phi^{(2)}$ are both nonlinear, with threshold behaviour at $\phi^{(i)}\sim0.35$. The curve, $\omega^{(1)}=\omega^{(1)}(\phi^{(1)})$ at $\phi^{(2)}=0$ is identical to that shown in figure 2 of \cite{tavoni2012}. Although the nonlinear factors of both terms in (\ref{eqn:ost}) are identically parameterised, $\omega^{(1)}$ depends more weakly on $\phi^{(2)}$ than $\phi^{(1)}$ because the second term is scaled by the weak social coupling. The ostracism components from the two communities add, producing effects not seen when ostracism is only local. Defectors in community $i$ experience nonzero ostracism even when $\phi^{(i)}$ is below its threshold, provided $\phi^{(j)}$ ($j\neq i$) is above its threshold ($\phi^{(i)}\lesssim0.35\lesssim\phi^{(j)}$). Moreover, when cooperators dominate both communities ($\phi^{(i)}\gtrsim0.5$), defectors in each experience a higher level of ostracism than can be applied by their local communities alone.  

The cooperative fraction of each subsystem's population evolves according to mean-field replicator dynamics \cite{hofbauer1998}. For subsystem 1, this is written as in \cite{tavoni2012} but using our dimensionless variables:
\beqn
{\dot{\phi}}^{(1)}(\tau)&=&\frac{R_m}{d}{\phi}^{(1)}(\tau)(1-{\phi}^{(1)}(\tau))\nonumber\\
&{}&\times\left(\frac{{\pi_d}^{(1)}(\tau)-{\pi_c}^{(1)}(\tau)}{{\pi_d}^{(1)}(\tau)}\right)(\omega^{(1)}(\tau)-{\pi_d}^{(1)}(\tau)).\label{eqn:dphi}
\eeqn

\subsection{Cases for modelling coupled system}
When $\delta=\lambda=0$, the original TSL model \cite{tavoni2012} describes each subsystem. In section \ref{sec:isosub}, we summarise its features as a benchmark for subsystem behaviour. In the sections below, we investigate how biophysical and social perturbations due to subsystem couplings shift subsystem fixed points, relative to an isolated subsystem, in different cases. First, we quantify how biophysical coupling affects each subsystem's norm of socially optimal harvesting when communities share resource information during norm setting. Two scenarios are obtained, in which norms respectively are and are not adjusted for coupling. We compute subsystem 1's biophysical and social fixed points for both scenarios, in cases where coupling is purely biophysical, purely social, or a combination of the two. We fix coupling strengths such that interactions between subsystems are at least twice as weak as interactions within them ($0\leq\delta\leq0.5$, ~$0\leq\lambda\leq0.5$) and also fix the cooperator-defector effort multiplier for subsystem 2, $\mu^{(2)}$ arbitrarily at $75\%$ of the range of $\mu$ values in an isolated subsystem ($\mu^{(2)}=0.75(\mu_N-1)+1=2.88$). This is because the range of $\mu$ depends on the biophysical coupling strength, so varies across the parameter space (see section \ref{sec:norms}). For consistency, therefore, $\mu^{(2)}$ is fixed relative to the isolated case.   

Prior analysis \cite{schluter2016} simulated the full coupled subsystem dynamics for each parameter combination. Initial conditions were chosen (${\phi^{(2)}}(0)$ always 0.9, ${\phi^{(1)}}(0)$ varied) and the system numerically evolved to equilibrium. Here, since we assume equilibrium and analyse the stability landscape, we instead fix the long-time behaviour of subsystem 2 and ignore its transient behaviour, which may originate anywhere in the chosen equilibrium's basin of attraction. For each scenario of norms and couplings, we consider cases in which subsystem 2 equilibrates at each of its monomorphic fixed points (${\phi^{(2)}}^*=0,1$). Although the stabilities of these fixed points vary across the parameter space, they are always accessible through an appropriate choice of initial conditions. This is not true for the mixed fixed points, which vary in both coordinates and stability. 

For each case, we address two questions: 1) is subsystem 1 resilient to cross-scale perturbations from subsystem 2 or do they cause it to undergo a stability phase transition; 2) does the long-time behaviour of subsystem 2 promote or prevent similar long-time behaviour in subsystem 1 and, therefore, the overall system (both subsystems together)?  
\section{RESULTS}

\subsection{Norm scenarios in coupled system} \label{sec:norms}
 In each subsystem, the effort level satisfying (\ref{eqn:Ec}) when the resource is at equilibrium quantifies the socially optimal harvesting norm. To compute subsystem 1's resource equilibria, we assume that both subsystems' zeroth- and first-order resource dynamics have equilibrated. Setting $\dot{\rho_1}^{(1)}(\tau)=\dot{\rho_1}^{(2)}(\tau)=0$ yields 
 
 \begin{subequations}
\beqn
{\rho_1^{(1)}}^*&=&\frac{d({\rho_0^{(2)}}^*-{\rho_0^{(1)}}^*)}{R_m {E^{(1)}}^*+2d{\rho_0^{(1)}}^*}\label{eqn:rho1equil0}\\
&=&\frac{R_m({E^{(1)}}^*-{E^{(2)}}^*)-S({E^{(1)}}^*)+S({E^{(2)}}^*)}{2 S({E^{(1)}}^*)}.~~~~~\label{eqn:rho1equil1}
\eeqn
\end{subequations}

When multiplied by $\delta$, this gives the shift in subsystem 1's resource equilibrium due to transfer between subsystems (relative to ${\rho_0^{(1)}}^*$). Substituting (\ref{eqn:rho1equil1}) and (\ref{eqn:rho0equil}) into (\ref{eqn:solnp}) gives

\begin{align}
{\rho^{(1)}}^*&\approx{\rho_0^{(1)}}^*+\delta{\rho_1^{(1)}}^*\nonumber\\
&\approx\frac{1}{2S({E^{(1)}}^*)d}\left(4cd+R_m {E^{(1)}}^*(R_m {E^{(1)}}^*-S({E^{(1)}}^*))\right.\nonumber\\
&{}~+\left.\delta d(R_m({E^{(1)}}^*-{E^{(2)}}^*)-S({E^{(1)}}^*)+S({E^{(2)}}^*))\right).\label{eqn:rhoshift}
\end{align}

Empirical studies have shown that communication leading to norm-based collaboration between common pool resource harvesters can strongly improve resource sustainability compared with cases of no communication \cite{ostrom2009}. We consider two scenarios for norm setting, called `uncollaborative' and `collaborative'. In the former, communities set norms without communicating, as if the subsystems were isolated ($E_c^*\equiv {E_c}_0^*=0.483$ and $E_N^*\equiv {E_N}_0^*=1.83$; $\mu_N={E_N}_0^*/{E_c}_0^*\equiv{\mu_N}_0=3.78$, as calculated in section \ref{sec:isosub}). In the collaborative scenario, communities perfectly communicate and set norms using resource information from both subsystems. Substituting (\ref{eqn:rhoshift}) into (\ref{eqn:prod}), and the result into (\ref{eqn:Ec}) and (\ref{eqn:En}), gives shifted $E_c^*$ and $E_N^*$ values. Numerically solving for these values across the weak coupling range ($0\leq\delta\leq0.5$) yields distributions that are very well approximated by linear fits from least-squares regression (Pearson correlation, $r>0.999$ for both): 
\beqn
E_c^*(\delta)={E_c}_0^*+7.70\times10^{-2}\delta\nonumber\\
\mu_N(\delta)={\mu_N}_0-5.56\times10^{-1}\delta.
\eeqn
In the collaborative scenario, these shifts must be included in both $\rho_0^{(i)}$ and $\rho_1^{(i)}$ to calculate $\rho^{(i)}$. 
  
\subsection{Purely biophysical coupling}
  
We first study cases in which $~0<\delta\leq0.5$, $\lambda=0$. Evaluating (\ref{eqn:rho1equil1}) determines shifts in biophysical equilibria across the parameter space (see fig. \ref{fig:resshift}). As (\ref{eqn:rho1equil0}) makes clear, ${\rho_1^{(1)}}^*$ varies in proportion to the difference in zeroth-order resource equilibria; when ${\rho_0^{(1)}}^*$ is higher than ${\rho_0^{(2)}}^*$, the first-order dynamics transfer resource from subsystem 1 to 2, and conversely. In turn, this difference depends on the social fixed point at which subsystem 2 is assumed to equilibrate. When ${\phi^{(2)}}^*=0$ (fig. \ref{fig:resshift}a,b), subsystem 2 is purely defective, the corresponding high level of extractive effort results in lower ${\rho_0^{(2)}}^*$, and the equilibrium first-order dynamics persistently transfer resource from subsystem 1 to 2. Conversely, when ${\phi^{(2)}}^*=1$  (fig. \ref{fig:resshift}c, d), the correspondingly low extractive effort gives higher ${\rho_0^{(2)}}^*$ and the first-order equilibrium persistently transfers resource from subsystem 2 to 1. 

Solving (\ref{eqn:dphi}) when $\dot{\phi}^{(1)}=0$ yields monomorphic social fixed points at ${\phi^{(1)}}^*=0,1$, and shifted mixed social fixed points on loci generally defined by $\omega^{(1)}({\phi^{(1)}}^*,{\phi^{(2)}}^*)=\pi_d^{(1)}(e_c,\mu^{(1)},{\phi^{(1)}}^*,{\rho^{(1)}}^*)$. More explicitly,
  
  \begin{dgroup}
 \begin{dmath*}
hn\left(e^{T e^{g {\phi^{(1)}}^*}}+\lambda e^{T e^{g {\phi^{(2)}}^*}}\right)\nonumber\\=e_c \mu^{(1)}\left(\gamma{\left[{E^{(1)}}^*\right]}^{\alpha-1} \left[\frac{R_m}{2Q({E^{(1)}}^*)} \nonumber \\\times\left(\delta\left(2({E^{(1)}}^*-{E^{(2)}}^*)-Q({E^{(1)}}^*)+Q({E^{(2)}}^*)\right)\nonumber\\~~~~~~~~~~~~+2Q({E^{(1)}}^*)\left(Q({E^{(1)}}^*)-2{E^{(1)}}^*\right)\right)\right]^\beta-w\right),\label{eqn:socshift}
\end{dmath*}
\end{dgroup}
  
where $Q({E^{(i)}}^*)=\sqrt{1+4{{E^{(i)}}^*}^2}$. Figure \ref{fig:mixedeqbp} plots $\omega^{(1)}({\phi^{(1)}}^*,{\phi^{(2)}}^*)$ and $\pi_d^{(1)}(e_c,\mu^{(1)},{\phi^{(1)}}^*,{\rho^{(1)}}^*)$, with curve intersections corresponding to mixed social fixed points because ${\dot{\phi}}^{(1)}(\tau)=0$ when $\omega^{(1)}(\tau)={\pi_d}^{(1)}(\tau)$ in (\ref{eqn:dphi}); the ostracism and defector payoff balance. 

\begin{figure}[ht]
\includegraphics[angle=0,width=0.95\columnwidth]{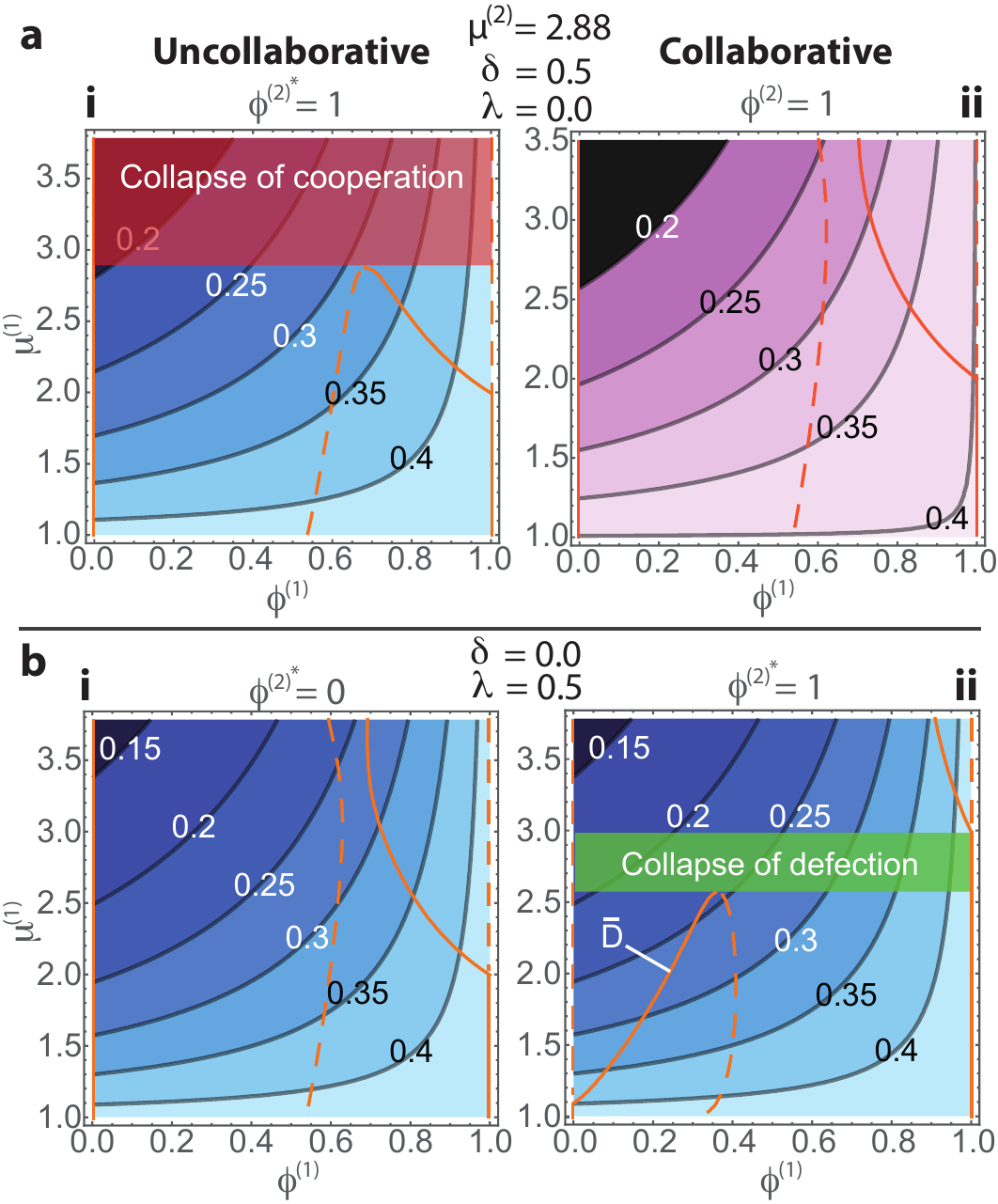}\centering
\caption{\textbf{a. Shifted biophysical (blue, purple) and social (orange) fixed points in subsystem 1 under purely biophysical, moderate ($\delta=0.5$) subsystem coupling.}  Results shown for the purely cooperative social equilibrium in subsystem 2 (${\phi^{(2)}}^*=1$), in the uncollaborative (panel i) and collaborative (ii) norm scenarios. Dashed and solid orange loci respectively comprise unstable and stable social fixed points. Defector payoff overwhelms ostracism in the red region in panel i. \textbf{b. Shifted subsystem 1 fixed points under purely social, moderate ($\lambda=0.5$) subsystem coupling.} Results shown for the purely defective (panel i) and purely cooperative (ii) social equilibrium in subsystem 2 (${\phi^{(2)}}^*=0,1$ respectively). Ostracism overwhelms defector payoff in the green region in panel ii. The previously unseen social attractor is labelled: $\bar{\text{D}}$ - mixed, defector dominated.}
\label{fig:ecocoup_soccoup}
\end{figure}

Figure \ref{fig:ecocoup_soccoup}a shows shifted biophysical and social fixed points in subsystem 1 under purely biophysical subsystem coupling. Results are shown for the upper bound of the weak coupling range ($\delta=0.5$, which we call moderate coupling), for both norm scenarios, when subsystem 2 has equilibrated at its purely cooperative social equilibrium (${\phi^{(2)}}^*=1$). When ${\phi^{(2)}}^*=0$ instead, subsystem 1 shows qualitatively similar behaviour to an isolated subsystem and is therefore resilient to the cross-scale biophysical perturbation (see fig. \ref{fig:ss2defectiveeqs}). However, figure \ref{fig:ecocoup_soccoup}ai shows a stability phase transition; when $\mu^{(1)}\gtrsim 2.9$, there are no mixed social fixed points because defector payoff exceeds ostracism for all values of $\phi^{(1)}$ (see fig. \ref{fig:mixedeqbp}b). Intuitively, this is because biophysical coupling gives subsystem 1 agents access to extra resource from subsystem 2, increasing their defector payoff, while without social coupling, subsystem 2 cannot provide extra ostracism to balance this payoff. Only the purely defective fixed point is stable in this region, so for any ${\phi^{(1)}}(0)<1$, collapse of cooperation is inevitable. This constitutes a CSI, since the weak perturbation from system-level dynamics generates a phase transition in subsystem 1, from multistable to monostable behaviour, and this cannot be predicted by analysing either subsystem in isolation. This is consistent with \cite{schluter2016}, in which this shift was described using agent-based methods\footnote{Due to small differences in model formulation, $\delta$ in \cite{schluter2016} (introduced in equation (4.1)) is equal to one quarter of $\delta$ here. e.g. $\delta = 0.5$ here is equivalent to $\delta=0.125$ in \cite{schluter2016}.}. Notably, cooperation never collapses in the collaborative norm scenario; operating under a norm set with the benefit of communication between subsystems preserves cooperation where it otherwise fails. Moreover, under both subsystem 2 equilibria (${\phi^{(2)}}^*=0,1$), the $\bar{C}$ equilibrium in subsystem 1 is more cooperative, for a given $\mu^{(1)}$ value, in the collaborative scenario. 

\subsection{Purely social coupling}\label{sec:soccoup}
We now consider cases in which $~0<\lambda\leq 0.5$ and $\delta=0$. Although one may question why communities that do not share resources would sanction each other for violating resource use norms, it is instructive to study such cases before combining social and biophysical coupling. When $\delta=0$, (\ref{eqn:solnp}) reduces to $\rho^{(1)}(\tau)=\rho_0^{(1)}(\tau)$, so the resource equilibrates as in an isolated subsystem and the uncollaborative and collaborative scenarios are identical. Moreover, the mixed social fixed points are independent of $\mu^{(2)}$ when $\delta=0$ in (\ref{eqn:socshift}). Figure \ref{fig:mixedeqs} shows the intersections of $\omega^{(1)}({\phi^{(1)}}^*,{\phi^{(2)}}^*)$ and $\pi_d^{(1)}(e_c,\mu^{(1)},{\phi^{(1)}}^*,{\rho^{(1)}}^*)$ (the mixed social fixed points) for purely social coupling.

Figure \ref{fig:ecocoup_soccoup}b shows how moderate purely social coupling ($\lambda=0.5,~\delta=0$) shifts system 1's fixed points for each of subsystem 2's monomorphic social equilibria. When ${\phi^{(2)}}^*=0$, subsystem 1's behaviour is qualitatively similar to an isolated subsystem and is therefore resilient to the cross-scale social perturbation. However, when ${\phi^{(2)}}^*=1$, subsystem 1 undergoes a stability phase transition. There is a mixed social equilibrium with a strong defector majority for $\mu^{(1)}\lesssim2.6$. For $\mu^{(1)}\gtrsim2.9$, there is a single mixed social equilibrium with a strong cooperator majority. In the intervening region ($2.6\lesssim\mu^{(1)}\lesssim2.9$), ostracism exceeds defector payoff for all values of ${\phi^{(1)}}$ and only the purely cooperative fixed point is stable. Intuitively, this is because social coupling exposes subsystem 1 defectors to additional ostracism from subsystem 2 cooperators, while the absence of biophysical coupling prevents subsystem 2 from increasing the subsystem 1 defector payoff through added resource availability. In this region, collapse of defection is inevitable in subsystem 1 for any ${\phi^{(1)}}(0)>0$. We see that the weak perturbation provided by social coupling between subsystems can produce two kinds of stability phase transitions at the subsystem level, from a multistable to a monostable phase and between two different multistable phases.

\subsection{Combined biophysical and social coupling}
\begin{figure}[ht]
\includegraphics[angle=0,width=0.97\columnwidth]{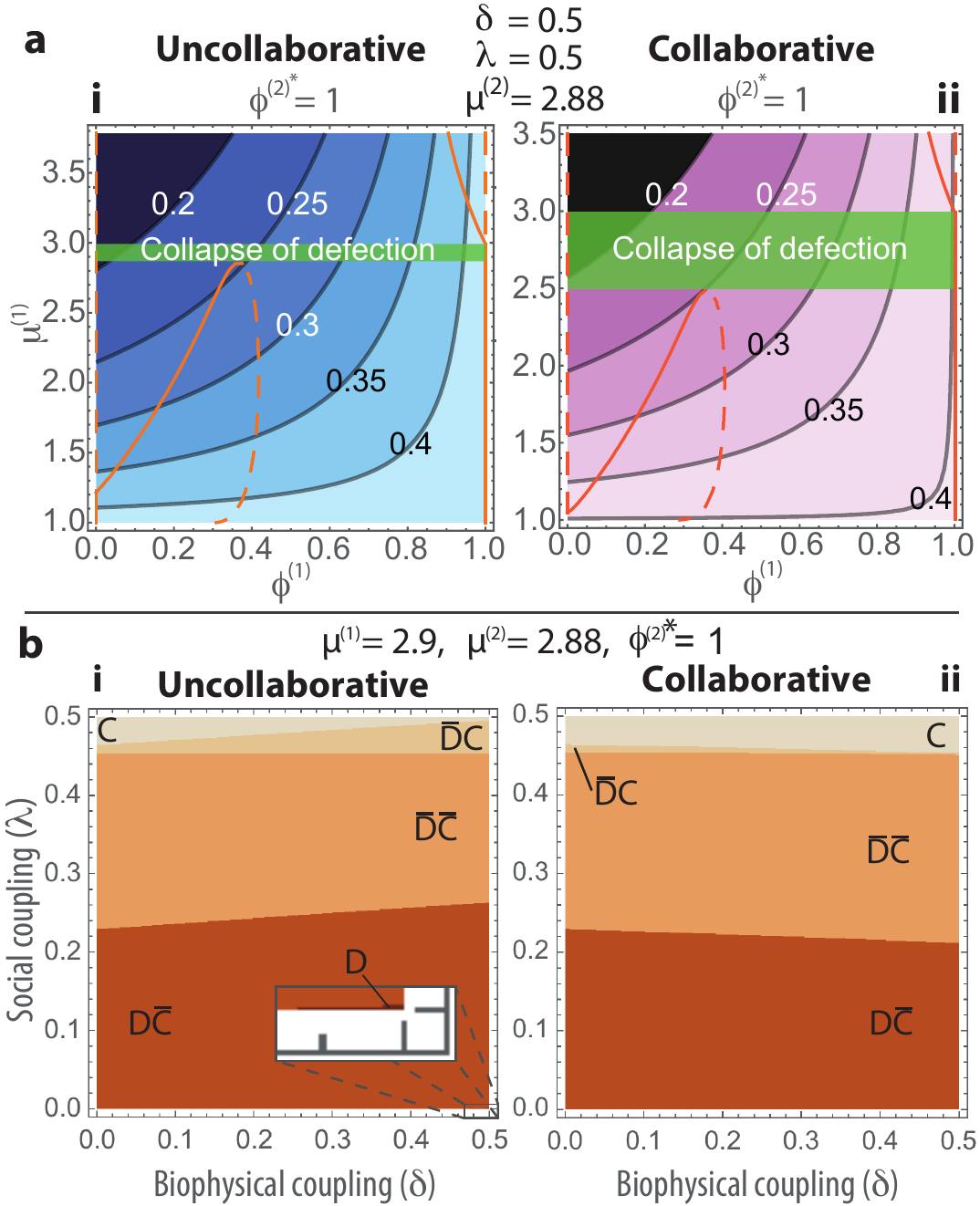}\centering
\caption{\textbf{a. Shifted biophysical (blue, purple) and social (orange) fixed points in subsystem 1 under combined biophysical and social subsystem coupling.} Results shown for equally moderate couplings ($\delta=\lambda=0.5$), for the purely  cooperative social equilibrium in subsystem 2 (${\phi^{(2)}}^*=1$). Ostracism overwhelms defector payoff in the green regions. \textbf{b. Phase plots showing how subsystem 1 social stability phase depends on social and biophysical coupling strengths, when $\mu^{(1)}=2.9$, $\mu^{(2)}=2.88$, ${\phi^{(2)}}^*=1$.} Results shown for 500 values, each, of $\delta$ and $\lambda$ ($2.5\times10^{5}$ combinations) per plot. Five stability phases are labelled, each comprising a landscape with either one or two of the four social attractor types described. The phase structure varies significantly across the ($\mu^{(1)},~\mu^{(2)},~{\phi^{(2)}}^*$) parameter space; values chosen here provide an illustrative example. \textbf{i.} Uncollaborative scenario, containing five phases. The D phase appears only in a small region at the lower right of panel ($\delta\gtrsim0.47$) and is magnified in the inset. \textbf{ii.} Collaborative scenario, containing four phases.}
\label{fig:ecosoccoup_phaseplots}
\end{figure}

Finally, we study interplay between weak biophysical and social couplings ($~0<\delta\leq 0.5$, $~0<\lambda\leq 0.5$). This represents a scenario in which two communities with a weak social relationship weakly share a common pool resource and enforce social norms within and between communities. Figure \ref{fig:mixedeqbps} shows the intersections of $\omega^{(1)}({\phi^{(1)}}^*,{\phi^{(2)}}^*)$ and $\pi_d^{(1)}(e_c,\mu^{(1)},{\phi^{(1)}}^*,{\rho^{(1)}}^*)$ (the mixed social fixed points) for equally moderate biophysical and social couplings ($\delta=\lambda=0.5$). Figure \ref{fig:ecosoccoup_phaseplots}a plots variations in subsystem 1's fixed points for subsystem 2's purely cooperative social equilibrium (${\phi^{(2)}}^*=1$). The collapse of subsystem 1 cooperation seen with purely biophysical coupling in the uncollaborative scenario (fig. \ref{fig:ecocoup_soccoup}ai) is prevented. Moreover, within small parameter ranges ($2.9\lesssim\mu^{(1)}\lesssim3$ in panel i and $2.5\lesssim\mu^{(1)}\lesssim3$ in ii), defection collapses instead. This shows that when biophysical and social coupling are equally moderate, and subsystem 2 equilibrates in a fully cooperative state, added ostracism experienced by subsystem 1 defectors overwhelms extra payoff available due to added resource access. The net result is a stability phase transition for subsystem 1, which does not occur without social coupling. Importantly, the long-time results of social interactions between subsystems overwhelm the effects of their biophysical interactions even when social and biophysical couplings are equal in strength. 

The results reveal important differences between norm scenarios. Subsystem cooperativities are much more mutually sensitive in the uncollaborative scenario. Weak perturbations from subsystem 2 can cause opposing types of social stability phase transitions in subsystem 1; collapse of cooperation when coupling is purely biophysical and collapse of defection when social coupling is present. Conversely, in the collaborative scenario, subsystems do not risk collapse of cooperation due to perturbations from weak biophysical coupling. Social coupling can cause collapse of defection also in this case, however. Furthermore, subsystem 1's $\bar{C}$ equilibrium is more cooperative in the collaborative scenario, as previously seen for purely biophysical coupling. 

Figure \ref{fig:ecosoccoup_phaseplots}b shows how subsystem 1's social stability landscape depends on social and biophysical coupling strengths across the weak-coupling parameter subspace. Five stability phases are labelled, each with a different combination of social attractors. There is clear asymmetry between how social coupling and biophysical coupling determine the stability phase. Whereas changes in social coupling strength cause phase transitions for every value of biophysical coupling in both scenarios, the inverse is not true; for most social coupling values, no change in biophysical coupling causes a phase transition. Panel i shows a phase plot for the uncollaborative scenario. The collapse of cooperation seen in figure \ref{fig:ecocoup_soccoup}a is restricted to a small parameter region at the lower right corner. At the opposite extreme, where social coupling is moderate, lies the purely cooperative phase and this extends across the range of weak biophysical couplings. The spectrum of other stability phases lie between these extremes. In section \ref{sec:stabphases}, we briefly describe mechanisms underlying the behaviour of each phase. The overall trend is that increasing the social coupling causes successive phase transitions in the subsystem 1 stability landscape, gradually increasing its cooperativity. The collaborative scenario (fig. \ref{fig:ecosoccoup_phaseplots}bii) shows the same broad features except that the D phase is absent, for reasons explained in section \ref{sec:soccoup}. Additionally, whereas the phase boundaries increase monotonically with biophysical coupling in the uncollaborative scenario, they decrease monotonically in the collaborative scenario. Thus, for a given level of resource sharing, communication during norm setting will allow lower levels of social coupling to cause stability phase transitions in subsystem 1. Collaborative norm setting is therefore an investment which can later increase the power of social pressure as a resource-sustaining mechanism. In sum, the results show the importance of collaboration between communities, both in establishing norms and enforcing those norms through social sanctioning, even when biophysical and social coupling are much weaker between subsystems than within each.

\section{DISCUSSION}\label{disc}
Our results show that social and biophysical CSIs can profoundly affect how interacting communities manage a common pool resource, and how their subsystem-level behaviours interact to produce system-level outcomes. Even weak social interactions between communities can produce long-time behaviour significantly more cooperative than that which eventuates in their absence. This is important for governance because it reveals a mechanism by which resource managers can help to ensure that subsystem-level efforts produce desired system-level outcomes. When coupling is purely biophysical and at least moderate in strength ($\delta\gtrsim0.5$), and norms are established without communication, pursuing socially optimal resource use is effectively a zero-sum game between communities. In this perverse situation, desire for system-level sustainability expressed through efforts at the subsystem level is self-defeating; success for the subsystem guarantees failure for the system. However, enforcing norms between communities resolves this because increased cooperation in one community encourages cooperation in the other, allowing them to cooperate as whole communities. This system-level cooperation depends strongly on the social coupling strength, even within the weak range. Strengthening the coupling produces successively higher levels of cooperation through transitions between stability phases (fig. \ref{fig:ecosoccoup_phaseplots}b). Moreover, communities that share information during norm setting protect themselves against collapse of cooperation at the subsystem level, and increase the power of intercommunity social pressure to increase cooperation at the system level.

Methodologically, our dual analytical/numerical approach offers a powerful complement to purely numerical, agent-based methods \cite{schluter2016}. In the dual approach, finding analytical expressions for subsystem fixed points enables their direct evaluation, circumventing the need for full dynamical simulation and drastically accelerating characterisation of the stability landscape across the parameter space. However, this comes at the cost of statistical behaviour outside the mean field, which is preserved by agent-based methods. There is also a small accuracy cost incurred by the perturbative biophysical model (see section \ref{sec:asspert}), though this can in principle be reduced by truncating (\ref{eqn:solnp}) at a higher order, notwithstanding associated increases in mathematical difficulty. In addition to efficiently characterising a two-level system, our dual approach is well suited to future study of larger systems, including more subsystem interaction mechanisms and/or more hierarchical levels. Powerful methods have been developed within the theory of dynamical systems and statistical physics, for analysing interplay between structure and dynamics at different scales in many-body systems. Singular perturbation theory \cite{kevorkian2012} and renormalisation \cite{mccomb2007}, for example, were first developed as refinements to the regular perturbation theory on which our biophysical model (\ref{eqn:drho0}, \ref{eqn:drho1}) is based. Renormalisation theory was refined into a technique for coarse-graining the statistical description of systems with many length and time scales, concisely explaining relationships between behaviour at different scales. This has provided deep insight into critical transitions in a variety of systems. Although it is acknowledged that dynamical systems methods are useful for modelling SESs, and that there is a need to better understand the multiscale behaviour of SESs and how this relates to their critical transitions, such multiscale methods have yet to be applied in this field. The formalism developed here is naturally suited to extension through such methods, which we suggest for future research. We foresee complementary roles for dual analytical/numerical dynamical systems methods and agent-based simulation, with the latter providing a computational laboratory in which to test the former in small prototypic systems before application to systems impractically large for full simulation.   

Even before extending the model to larger systems, there are many options for refinement. One is to modify the social interaction model by applying results from psychology. Research on in-group bias suggests that members of a community (in-group) are likely to sanction members of another community (out-group) more harshly than in-group members for norm violation \cite{mcauliffe2016}. This could be incorporated into our model as dependence of the parameters controlling the curvature and maximum of the ostracism experienced by out-group defectors ($h, T$ and $g$ in the second term of (\ref{eqn:ost})) on the social coupling between communities ($\lambda$). Furthermore, psychological limits to human social network size and structure \cite{sutcliffe2012} could be incorporated by limiting the social connectedness available to each agent and, therefore, their capacity to ostracise and be ostracised by others. Other interesting results may come from breaking the parametric symmetries assumed to exist between subsystems in this study. Ultimately, more realistic representations of SESs will come from using a refined subsystem model as a building block in larger, multilevel, multiscale systems with parameterisations statistically distributed across subsystems. The multiscale analysis methods described above are well suited to such a system. Used in combination with statistical and qualitative analyses from multiscale empirical studies, such modelling methods offer strong potential for better understanding and managing CSIs in SESs.

 Our work raises fundamental questions about what is meant by 'scale' in SESs, and how best to model it. Different usage of the term across natural and social sciences influencing SES research has produced tension in the literature \cite{cash2006,gibson2000}, which must be navigated if a coherent theory of CSIs is to be developed. Natural sciences usually define scales as orders of magnitude in spatial and temporal separations. Social sciences typically define scales as levels of organisation in hierarchical social systems, though there has also been interest in how spatial embeddedness affects social systems \cite{logan2012, butts2012, daraganova2012}. In modelling studies, CSIs are commonly posed as interactions between quantities which vary over different spatial and temporal scales but precisely how such interactions depend on space and time is often not considered. In this study, we considered interactions across scales in the strengths of biophysical and social couplings within and between subsystems. One possibility for interpreting our results in terms of conventional spatiotemporal framing of CSIs is to determine how biophysical and social couplings depend on space and time. However, given the vast range of resource pool types and social systems which may be studied, this is a nontrivial undertaking. Intuitively, one may expect coupling strengths to vary inversely with spatiotemporal separation and this may indeed be the case in some systems. However, in a globalised, digitally connected world, couplings can depend counterintuitively on, or be effectively independent from, spatiotemporal separations, as is the case for so-called teleconnections \cite{scholes2013}. Recent work on social-ecological network analysis \cite{sayles2017, baggio2016, bodin2012} has begun to quantify how the spatiotemporal embeddedness of social and biophysical dynamics affects SES resilience. We propose that further research in this direction, across a broad range of systems and contexts, may be an important step towards a more general and broadly applicable theory of CSIs.


\section{AUTHORS' CONTRIBUTIONS}
All authors jointly designed the research. AKR developed the analysis and interpreted the results under advice from SJL and MS. AKR wrote, and SJL and MS critically revised, the manuscript. All authors gave final approval for publication. 

\section{ACKNOWLEDGEMENTS}
AKR thanks Alessandro Tavoni, Sonja Radosavljevic and Supriya Krishnamurthy for helpful discussions.

\section{FUNDING}
This project received funding from the European Research Council (ERC) under the European Union’s Horizon 2020 research and innovation programme (grant agreement No 682472 — MUSES), and Swedish Research Council Formas project grant 2014-589.

\bibliographystyle{9}

\clearpage

\beginsupplement
\section{SUPPORTING INFORMATION}\label{sec:SI}

\subsection{Isolated subsystem: original TSL model}\label{sec:isosub}
When $\delta=\lambda=0$, the original TSL model \cite{tavoni2012} describes each subsystem. Dropping labels for brevity, we summarise its features as a benchmark for assessing effects of perturbations due to biophysical and social couplings in the coupled system. TSL calculated biophysical and social fixed points for a range of initial social conditions, $\phi(0)$ and parameters, $\mu$. Setting $\dot{\rho}(\tau)=0$, (\ref{eqn:drho0}) gives biophysical fixed points as a function of total community effort:

\be
\rho_0^*=\frac{q}{2 d}\left(-E^* R_m+S(E^*)\right),\label{eqn:rho0equil}
\ee

where $S(E^*)=\sqrt{4 c d+{E^*}^2 R_m^2}$ and asterisks denote fixed-point values. Substitution into (\ref{eqn:prod}) gives socially optimal (cooperative) and individually optimal (Nash) effort levels ($E_c^*,~E_N^*$), conditions for which are \cite{dasgupta1979} respectively 
\begin{subequations}
\beqn
w&=&\frac{df}{dE^*}~~\text{and}\label{eqn:Ec}\\ w&=&\frac{f}{E^*}-\frac{1}{n}\left(\frac{f}{E^*}-\frac{df}{dE^*}\right).\label{eqn:En} 
\eeqn
\end{subequations}
It is assumed that $1<\mu\leq \mu_N$, where $\mu_N=E_N^*/{E_c^*}$. Using parameters from \cite{tavoni2012} ($n=50,~q=1,~c=50,~d=50,~R_m=200,~\alpha=0.6,~\beta=0.2,~\gamma=10$) gives $E_c^*=0.483,~E_N^*=1.83$ and corresponding resource fixed points $\rho_{0_c}^*=0.425,~\rho_{0_N}^*=0.134$. Socially optimal and Nash efforts for individual agents are $e_c=E_c^*/n=9.65\times10^{-3}$ and $e_N=E_N^*/n=3.65\times10^{-2}$. The former quantifies the norm of socially optimal harvesting. It is assumed that $e_c<e_d\leq e_N$, so $1<\mu\leq3.78$, since $e_N/e_c\equiv\mu_N=3.78$. 
Setting $\dot{\phi}(\tau)=0$ in (\ref{eqn:dphi}) yields monomorphic social fixed points ($\phi^*=0,1$) and mixed social fixed points on loci described by 
$\omega(\phi^*)=\pi_d(e_c,\mu,\phi^*,\rho_0^*)$.

\begin{figure}[ht]
\includegraphics[angle=0,width=0.6\columnwidth]{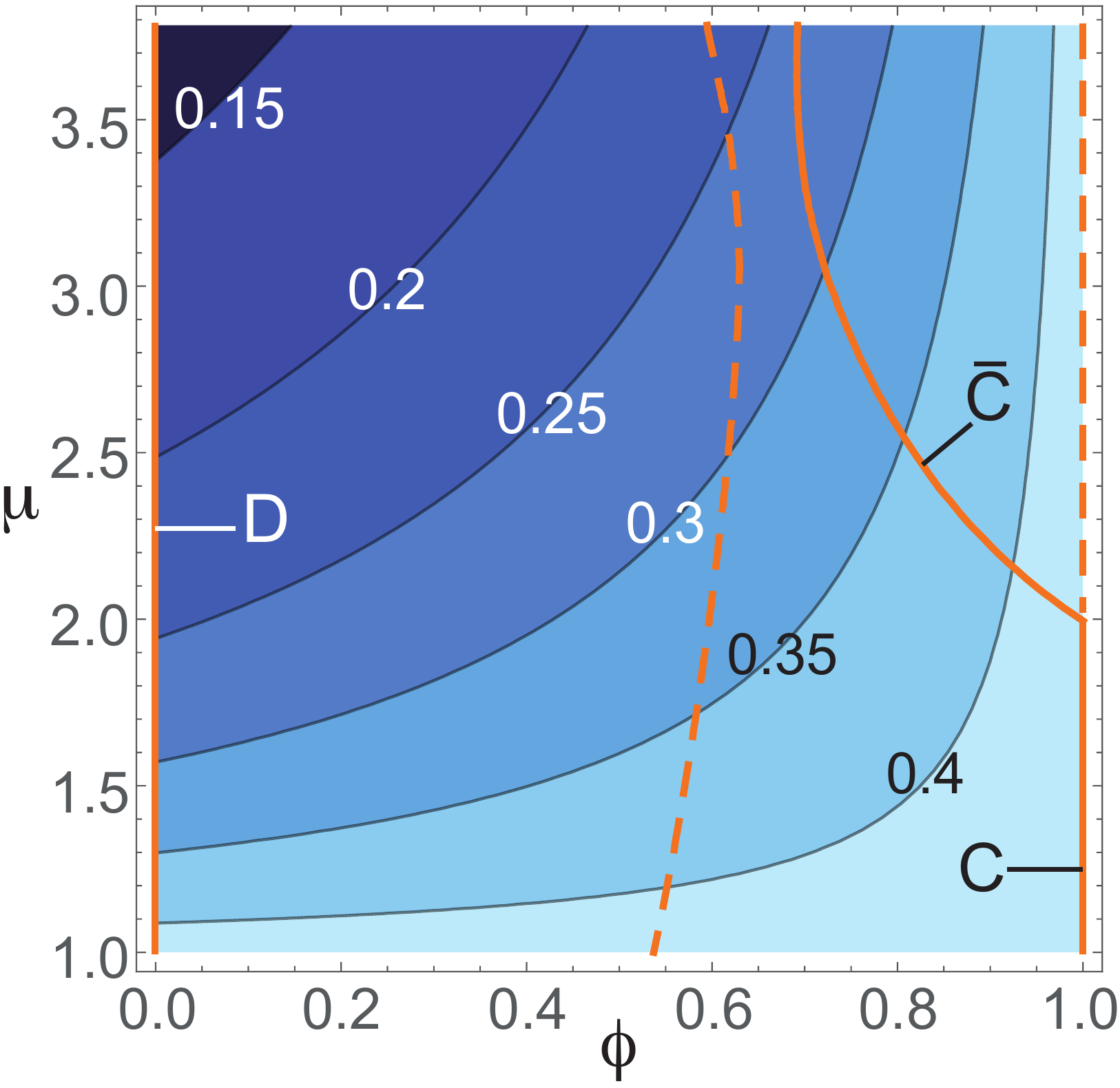}\centering
\caption{\textbf{Social (${\phi}^*$ -- orange) and biophysical (${\rho}^*$ -- blue) fixed points of an isolated TSL subsystem, parameterised by cooperative population fraction ($\phi(\tau)$) and defector-cooperator effort multiplier ($\mu$).} Dashed and solid orange loci respectively comprise unstable social fixed points and stable social fixed points (equilibria). The equilibria are labelled by type: D- purely defective, C- purely cooperative, $\bar{\text{C}}$- mixed, cooperator dominated. Parameters: $s=6.8\times10^{-3},~T=-150,~g=-10$. See figure 3 in \cite{tavoni2012} for more information.}
\label{fig:uncoupequil}
\end{figure}

Figure \ref{fig:uncoupequil} overlays these on corresponding biophysical fixed points (\ref{eqn:rho0equil}). The stable fixed points (equilibria) are attractors toward which the system evolves. The $\mu$ parameter space is divided into two social stability phases (qualitatively different stability landscapes, each persisting across some parameter range\footnote{A familiar example of phases comes from thermodynamics, in which substances such as water show the phases, solid, liquid and gas across different ranges of temperature and pressure.}), characterised by different combinations of attractors. When $\mu\gtrsim2.0$, there are two social attractors, one purely defective ($\phi^*=0$) and one mixed and cooperator-dominated ($0.5<\phi^*<1$). We label these attractor types, respectively, D and $\bar{\text{C}}$. When $\mu\lesssim2.0$, D and a purely cooperative attractor (C; $\phi^*=1$) are present.

\subsection{Assessment of perturbative approximation} \label{sec:asspert}
\begin{figure*}[ht]
\includegraphics[angle=0,width=0.7\textwidth]{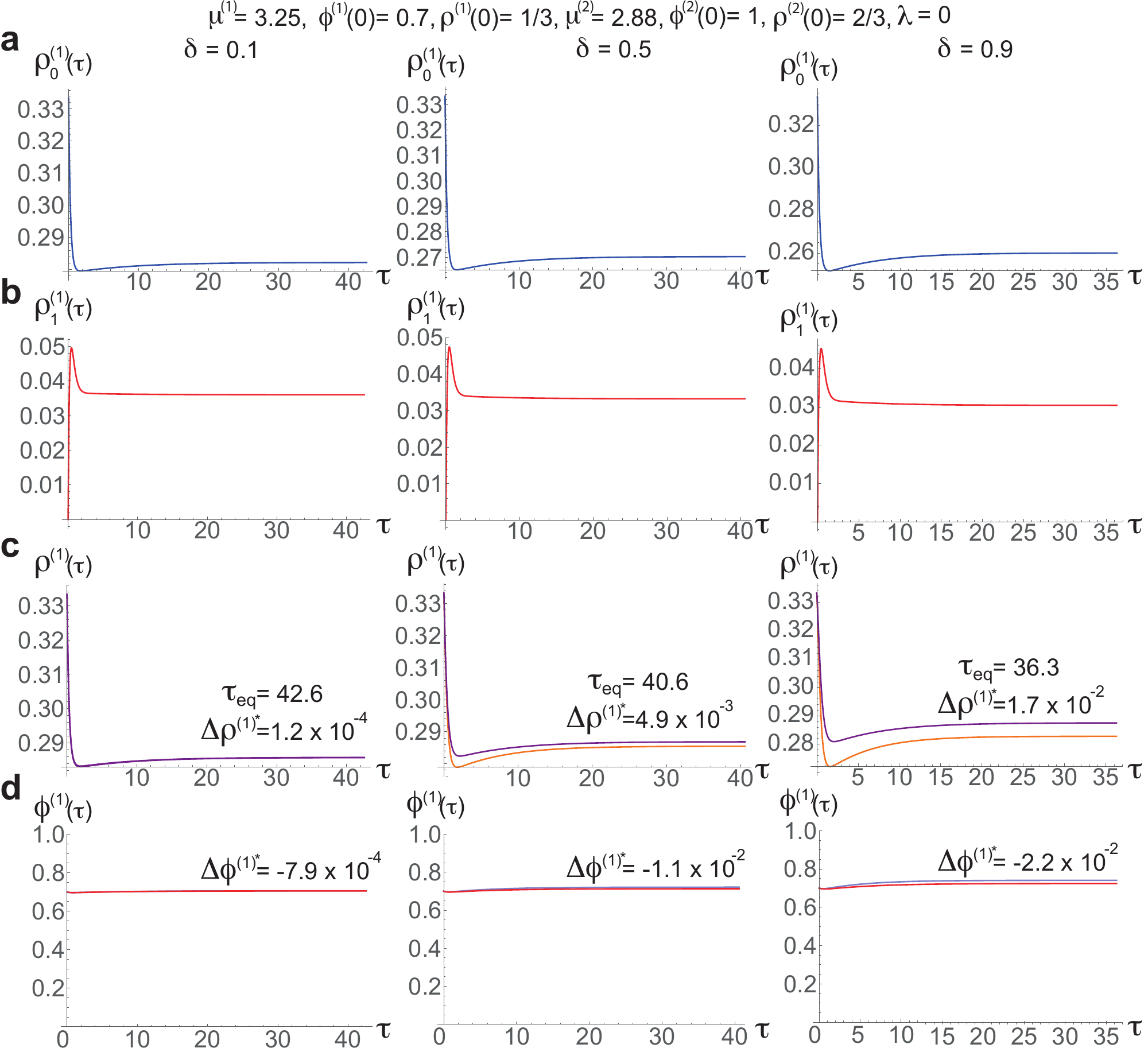}\centering
\caption{\textbf{Comparison of biophysical and social dynamics of subsystem 1, in the presence of purely biophysical subsystem coupling, between full numerical solution and perturbative approximation of the biophysical dynamics.} Cases of very weak (\textbf{left}), weak (\textbf{middle}) and strong (\textbf{right}) coupling are shown. \textbf{a}: zeroth-order resource dynamics, ${\rho_0^{(1)}(\tau)}$; \textbf{b}: first-order resource perturbation dynamics, ${\rho_1^{(1)}(\tau)}$; \textbf{c}: comparison of perturbative approximation $\left(\rho^{(1)}(\tau)\approx\rho_0^{(1)}(\tau)+\delta\rho_1^{(1)}(\tau),~\text{purple}\right)$ with full numerical resource dynamics (orange); \textbf{d}: comparison of cooperative population fraction dynamics between full numerical solution (blue) and solution using perturbative approximation for the biophysical dynamics (red). Parameters and initial conditions shown at top.}
\label{fig:transient}
\end{figure*}

We approximate the biophysical dynamics in the coupled subsystem model (\ref{eqn:drho}) with a perturbation expansion (\ref{eqn:solnp}), truncated at first order. Here we assess this approximation by comparing it with the full numerical solution of the mean-field replicator dynamics (\ref{eqn:drho}). Figure \ref{fig:transient} compares both the transient dynamics and equilibria for a range of coupling strengths. In each case, calculations were run until the system converged sufficiently close to equilibrium that the following condition was met:

\be
|\dot{\rho}^{(1)}(\tau)|+|\dot{\rho}^{(2)}(\tau)|+|\dot{\phi}^{(1)}(\tau)|+|\dot{\phi}^{(2)}(\tau)|<\Gamma,\label{eqn:tol}
\ee

where $\Gamma$ is a tolerance which we set to $\Gamma=10^{-5}$, since this gave a good balance between accuracy and practicality. 

We label the lowest time at which (\ref{eqn:tol}) is satisfied $\tau_{eq}$, and define the relative error in the resource equilibrium obtained from the perturbative approximation,

\be
\Delta{\rho^{(1)}}^*=\frac{\rho_0^{(1)}(\tau_{eq})+\delta\rho_1^{(1)}(\tau_{eq})-\rho^{(1)}(\tau_{eq})}{\rho^{(1)}(\tau_{eq})},\label{eqn:rhoerror}
\ee

where $\rho^{(1)}(\tau_{eq})$ is the numerical solution of the full resource dynamics (\ref{eqn:drho}). Similarly, we define the relative error in the equilibrium cooperative fraction of the subsystem 1 population,

\be
\Delta{\phi^{(1)}}^*=\frac{\phi_p^{(1)}(\tau_{eq})-\phi^{(1)}(\tau_{eq})}{\phi^{(1)}(\tau_{eq})},\label{eqn:phierror}
\ee

where $\phi_p^{(1)}(\tau_{eq})$ is the cooperative fraction obtained when the perturbative approximation is used for the resource dynamics and $\phi^{(1)}(\tau_{eq})$ is the full numerical solution. These relative errors are assessed for each case in figure \ref{fig:transient}. The results show excellent agreement between the perturbative and full solutions in the cases of very weak ($\Delta{\rho^{(1)}}^*=1.2\times10^{-4},~\Delta{\phi^{(1)}}^*=-7.9\times10^{-4}$) and weak coupling ($\Delta{\rho^{(1)}}^*=4.9\times10^{-3},~\Delta{\phi^{(1)}}^*=-1.1\times10^{-2}$). Surprisingly, even though the strong coupling case (panel c) violates the weak coupling assumption underpinning our perturbative treatment ($0<\delta\ll1$), the relative errors in equilibrium resource  and cooperative population levels are both only $\sim2\%$ in this case. Although this varies across the parameter space, we found comparably small errors across a wide range of parameter values, suggesting that the perturbative treatment is more robust under strong biophysical coupling than might be expected. We leave elucidation of the reasons for this unexpected robustness to future work.

\subsection{Supplementary results}
\begin{figure}[ht]
\includegraphics[angle=0,width=0.95\columnwidth]{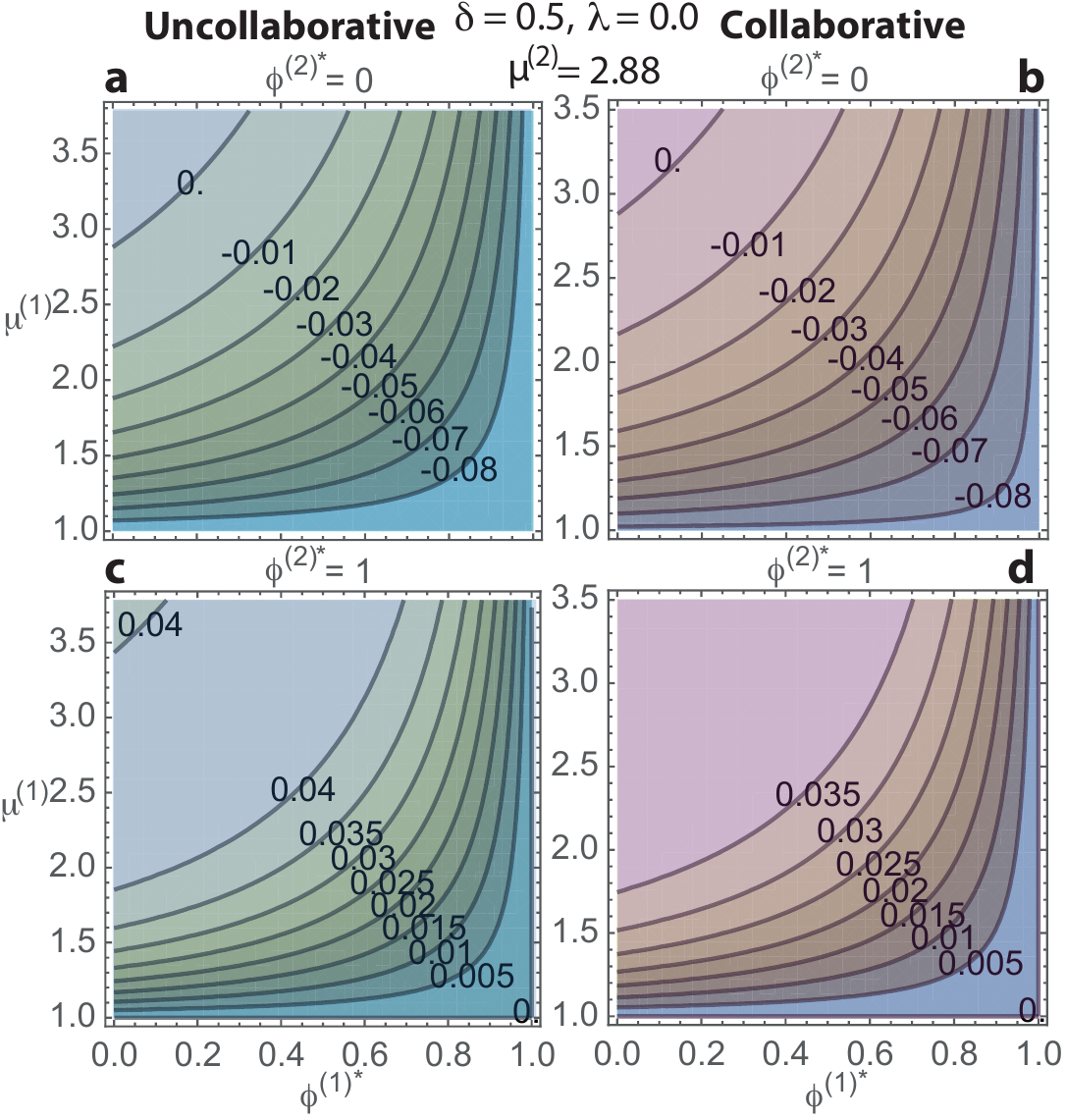}\centering
\caption{\textbf{Fixed points of the first-order biophysical dynamics in subsystem 1 (${\rho_1^{(1)}}^*$), in the uncollaborative and collaborative norm scenarios.} Per (main text eq'n \ref{eqn:solnp}), when multiplied by $\delta$ these give the biophysical coupling-induced shifts in the subsystem 1 fixed points compared with an isolated subsystem (see main text fig. \ref{fig:uncoupequil}). Results are shown for cases in which subsystem 2 equilibrates at each of its two monomorphic social fixed points. Values in panels a and b are negative because in those cases the equilibria correspond to ongoing transfer of resource from subsystem 1 to 2. The reverse is true for panels c and d.}
\label{fig:resshift}
\end{figure}

Here we present detailed results to explain the deviations in the subsystem 1 fixed points under different cases of coupling (main text fig's \ref{fig:ecocoup_soccoup}, \ref{fig:ecosoccoup_phaseplots}a), compared with an isolated subsystem (main text fig. \ref{fig:uncoupequil}). 
Evaluating (main text eq'n \ref{eqn:rho1equil1}) determines the shifts in biophysical equilibria across the parameter space (See fig. \ref{fig:resshift}). As (main text eq'n \ref{eqn:rho1equil0}) makes clear, ${\rho_1^{(1)}}^*$ varies proportionally to the difference in zeroth-order resource equilibria; when ${\rho_0^{(1)}}^*$ is higher than ${\rho_0^{(2)}}^*$, the first-order dynamics transfer resource from subsystem 1 to 2, and conversely.

\begin{figure}[ht]
\includegraphics[angle=0,width=\columnwidth]{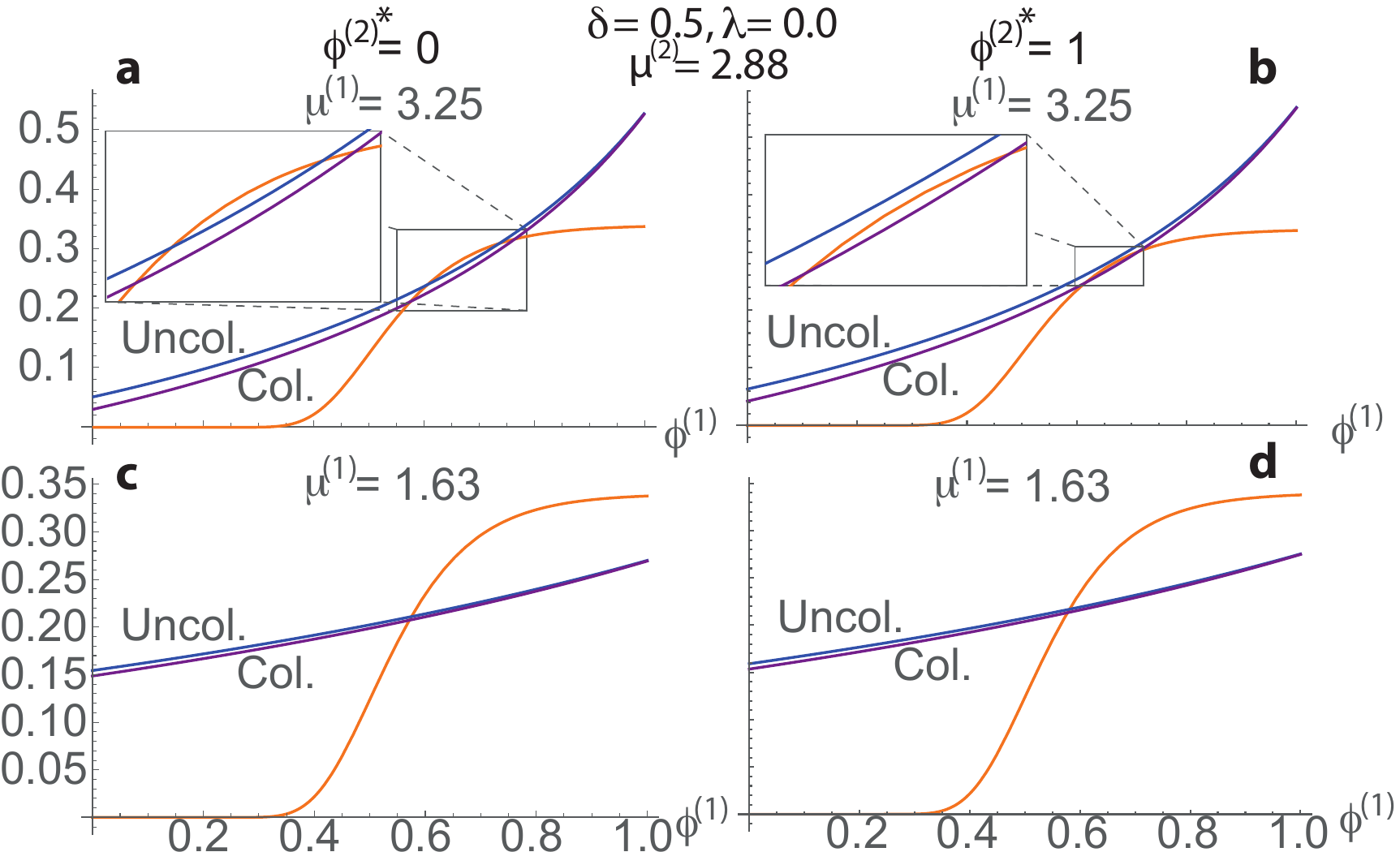}\centering
\caption{\textbf{Determination of mixed social fixed points in subsystem 1 in the case of purely biophysical subsystem coupling, in the uncollaborative and collaborative norm scenarios.} Fixed points are determined by intersections between the defector payoff (blue, purple) and ostracism (orange). Results shown for both monomorphic social equilibria in subsystem 2. In the uncollaborative scenario in b, there is no intersection, resulting in the collapse of cooperation.}
\label{fig:mixedeqbp}
\end{figure}

Figure \ref{fig:mixedeqbp} plots the ostracism ($\omega^{(1)}({\phi^{(1)}}^*,{\phi^{(2)}}^*)$) and payoff ($\pi_d^{(1)}(e_c,\mu^{(1)},{\phi^{(1)}}^*,{\rho^{(1)}}^*)$) experienced by subsystem 1 defectors when coupling is purely biophysical. The payoff differs between the uncollaborative and collaborative scenarios, and is shown for both. Curve intersections correspond to mixed social fixed points because ${\dot{\phi}}^{(1)}(\tau)=0$ when $\omega^{(1)}(\tau)={\pi_d}^{(1)}(\tau)$ in (\ref{eqn:dphi}); the ostracism and payoff experienced by defectors balance under this condition. When defectors harvest with effort only modestly higher than cooperators (e.g. 25\% higher: $\mu^{(1)}=0.25(\mu_N-1)+1=1.63$; panels c and d) there is only one mixed fixed point (unstable) and it is not significantly affected by either subsystem 2 cooperativity (${\phi^{(2)}}^*=0,1$) or norm scenario. However, when defectors harvest with effort close to the Nash level (e.g. $\mu^{(1)}=0.9(\mu_N-1)+1=3.25$), both of these quantities strongly affect subsystem 1 behaviour. There are two mixed fixed points in the collaborative scenario under both subsystem 2 cooperativity conditions, and also in the uncollaborative scenario when subsystem 2 is purely defective (${\phi^{(2)}}^*=0$). However, when ${\phi^{(2)}}^*=1$ (panel b), subsystem 1 has no mixed fixed points because the defector payoff, $\pi_d^{(1)}$, is higher than the ostracism, $\omega^{(1)}$, for all values of $\phi^{(1)}$. Intuitively, this is because biophysical coupling provides subsystem 1 agents access to additional resource from subsystem 2, which increases the defector payoff, while in the absence of social coupling, subsystem 2 cannot provide any extra ostracism to balance this increased payoff for subsystem 1 defectors.

\begin{figure}[ht]
\includegraphics[angle=0,width=\columnwidth]{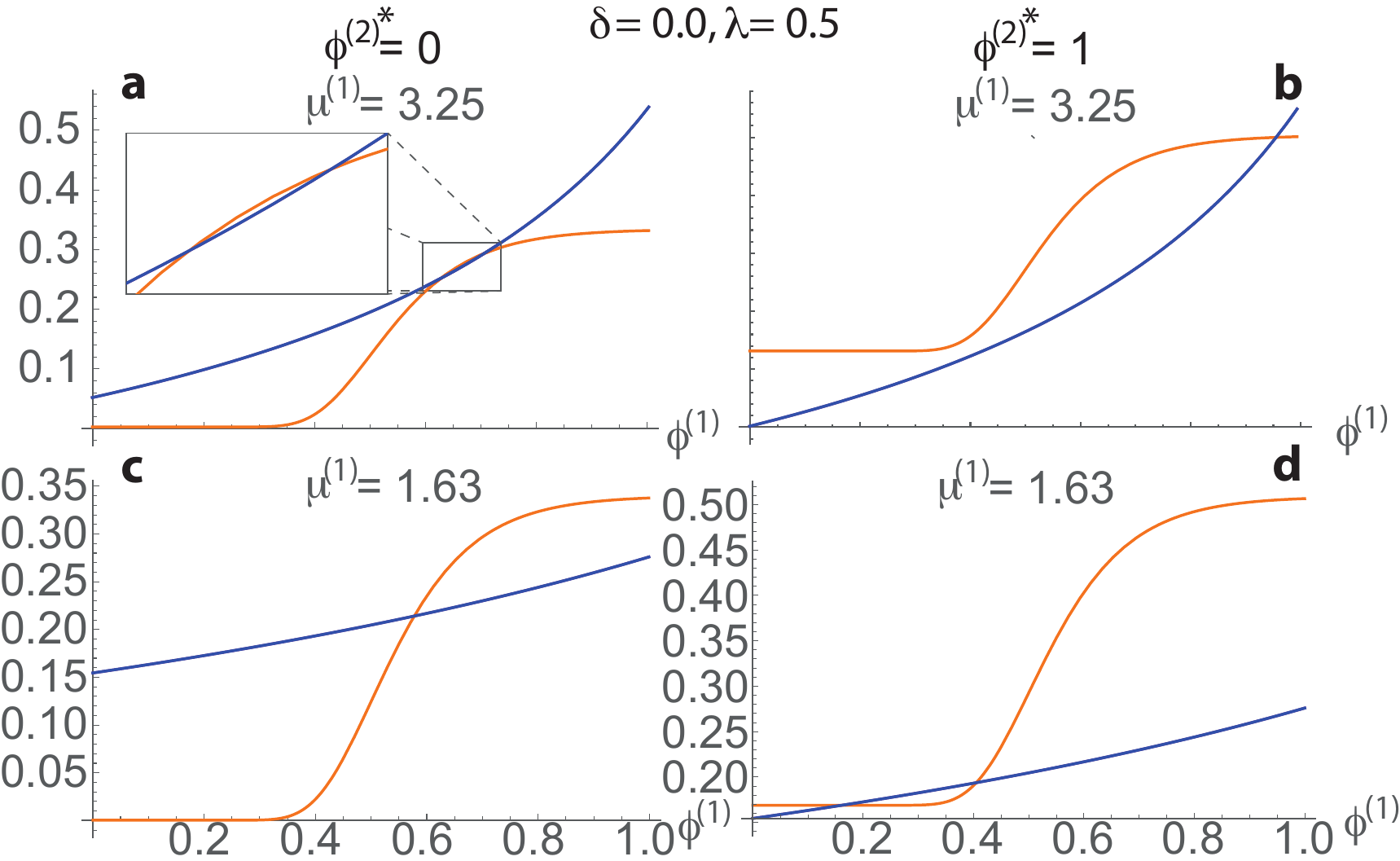}\centering
\caption{\textbf{Determination of mixed social fixed points in subsystem 1 in the case of purely social subsystem coupling, in the uncollaborative and collaborative norm scenarios (which are identical in this case).} Results shown for both monomorphic social equilibria in subsystem 2. Ostracism is nonzero even for small $\phi^{(1)}$ in panels b and d because ${\phi^{(2)}}^*=1$ and the social coupling conveys ostracism to subsystem 1 even when its own population is largely or completely defective. This enables the appearance of a fixed point in d, not seen in the absence of social coupling. This is a stable, mixed fixed point with a large defector majority.}
\label{fig:mixedeqs}
\end{figure}

Figure \ref{fig:mixedeqs} plots $\omega^{(1)}({\phi^{(1)}}^*,{\phi^{(2)}}^*)$ and $\pi_d^{(1)}(e_c,\mu^{(1)},{\phi^{(1)}}^*,{\rho^{(1)}}^*)$ in the case of purely social coupling. When subsystem 2 equilibrates at ${\phi^{(2)}}^*=0$ (panels a, c), the ostracism felt by defectors in subsystem 1 is significant only when ${\phi^{(1)}}^*$ reaches the threshold ($\sim0.35$) determined by the parameterisation of (\ref{eqn:ost}) (see fig. \ref{fig:uncoupequil}). Subsystem 1 behaviour then shows no phase transition in stability landscape relative to the isolated case; when defectors harvest only modestly harder than cooperators (e.g. $\mu^{(1)}=1.63$) there is a single mixed (unstable) fixed point (panel c), and when defectors harvest at close to the Nash effort ($\mu^{(1)}=3.25$), a second mixed (stable) fixed point occurs (panel a). However, when subsystem 2 equilibrates at its purely cooperative fixed point (${\phi^{(2)}}^*=1$; panels b, d), subsystem 1 defectors experience nonzero ostracism even when the cooperative population fraction in subsystem 1 is below its threshold ($\sim0.35$), due to the influence of subsystem 2 cooperators mediated by the social coupling. In this case, when $\mu^{(1)}=3.25$, only a single mixed (stable) social fixed point is observed, with a large cooperator majority (panel b). When $\mu^{(1)}=1.63$, however, two mixed social fixed points are present (one stable, one unstable), including one (stable) not seen in the case of purely biophysical coupling, with a large defector majority (panel d). 

\begin{figure}[ht]
\includegraphics[angle=0,width=\columnwidth]{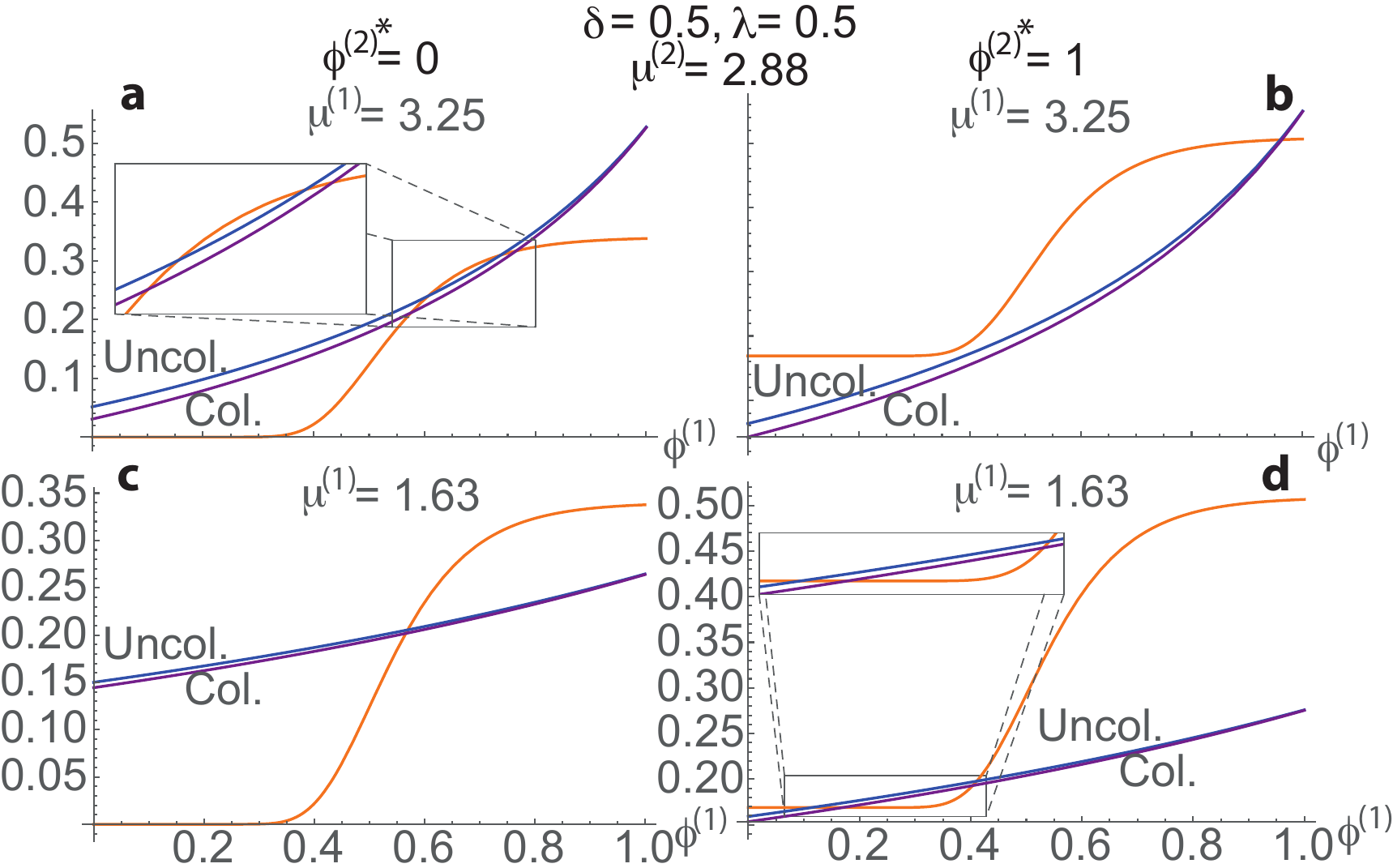}\centering
\caption{\textbf{Determination of mixed social fixed points in subsystem 1 in the case of combined biophysical and social subsystem coupling, in the uncollaborative and collaborative norm scenarios.} Results shown for both monomorphic social equilibria in subsystem 2. The norm scenarios differ only through small shifts in the fixed points and have no qualitative differences in the cases shown.}
\label{fig:mixedeqbps}
\end{figure}

\begin{figure}[ht]
\includegraphics[angle=0,width=0.95\columnwidth]{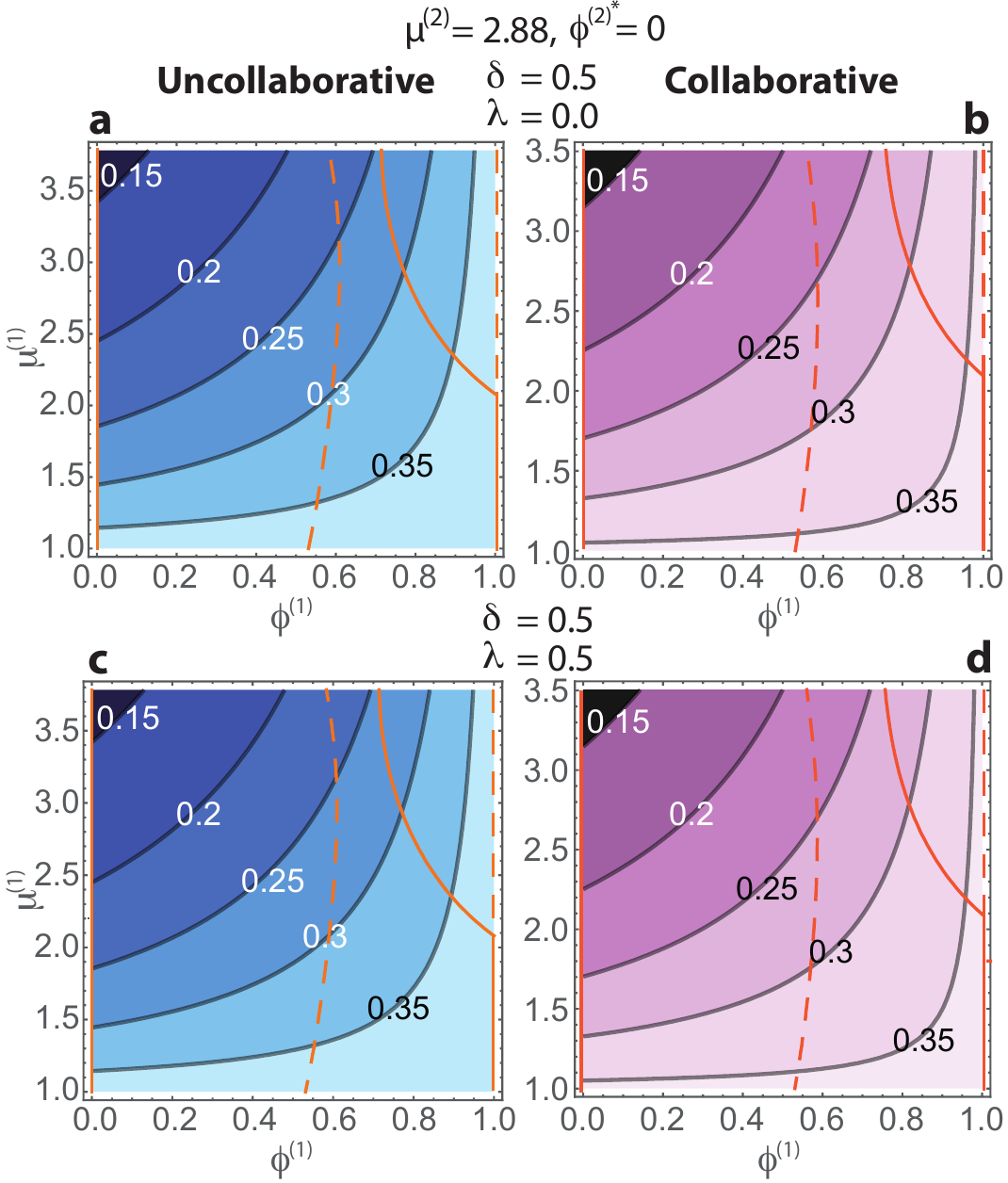}\centering
\caption{\textbf{Shifted biophysical (blue, purple) and social (orange) fixed points in subsystem 1 under purely biophysical subsystem coupling (panels a,b) and mixed social and biophysical coupling (c,d) when subsystem 2 equilibrates at its purely defective fixed point (${\phi^{(2)}}^*=0$)}. Results shown for uncollaborative (a,c) and collaborative (b,d) norm scenarios and moderate coupling ($\delta=0.5$, $\lambda=0.5$). Dashed and solid orange loci respectively comprise unstable and stable social fixed points. The results show only small deviations and no stability phase transitions compared with an isolated subsystem (fig. \ref{fig:uncoupequil}). The effect of subsystem 2 long-time behaviour (${\phi^{(2)}}^*=0$ vs. ${\phi^{(2)}}^*=1$) on subsystem 1 long-time behaviour under these couplings can be seen by comparing panels a,b of this figure with a-i and a-ii in figure \ref{fig:ecocoup_soccoup}, and panels c,d here with a-i and a-ii in figure \ref{fig:ecosoccoup_phaseplots}}
\label{fig:ss2defectiveeqs}
\end{figure}

Figure \ref{fig:mixedeqbps} plots $\omega^{(1)}({\phi^{(1)}}^*,{\phi^{(2)}}^*)$ and $\pi_d^{(1)}(e_c,\mu^{(1)},{\phi^{(1)}}^*,{\rho^{(1)}}^*)$, showing their intersections, when biophysical and social couplings are both moderate ($\delta=\lambda=0.5$). The results can largely be understood as a straightforward combination of the effects seen above, in figures \ref{fig:mixedeqbp} and \ref{fig:mixedeqs}. Of particular note, however, is the prevention of collapse of cooperation, in the uncollaborative scenario, previously seen for purely biophysical coupling (fig. \ref{fig:mixedeqbp}b), here replaced by a single highly cooperative equilibrium (\ref{fig:mixedeqbps}b). This occurs because the added ostracism due to social coupling overwhelms the added defector payoff due to extra resource availability from biophysical coupling. The same explanation underlies the replacement of the bistable behaviour previously seen in the collaborative scenario, under the same parameters, (fig. \ref{fig:mixedeqbp}b) also by a single highly cooperative equilibrium (fig. \ref{fig:mixedeqbps}b).

\subsection{Mechanisms underlying stability phases}\label{sec:stabphases}
Figure \ref{fig:ecosoccoup_phaseplots}b presents social stability phase plots for subsystem 1, showing up to five different phases across the weak coupling parameter space. In this section, we briefly summarise mechanisms underlying the behaviour of each phase. In the purely defective D region, additional ostracism due to very weak social coupling is inadequate to balance the extra defector payoff due to added resource availability from moderate biophysical coupling. Defector payoff exceeds ostracism for all subsystem 1 population compositions, making collapse of cooperation inevitable. With a very small increase in social coupling we shift to the D$\bar{\text{C}}$ phase. Here, added ostracism from subsystem 2 enables the ostracism experienced by subsystem 1 defectors to balance the defector payoff but only when ostracism from subsystem 1 is also high, which occurs only with a cooperative majority. This phase includes the isolated subsystem case ($\delta=\lambda=0$) and region of what may be considered very weak couplings ($\delta\lesssim0.1,~\lambda\lesssim0.1$). Increasing social coupling again, we shift to the $\bar{\text{D}}\bar{\text{C}}$ phase, in which added ostracism from subsystem 2 is sufficient to balance the defector payoff even when subsystem 1 has a defector majority, giving rise to the mixed, defector-dominated attractor. The mixed, cooperator-dominated attractor persists in this phase because the ostracism and defector payoff can also balance when subsystem 1 has a cooperator majority (see e.g. fig. \ref{fig:mixedeqbps}a, b). With yet higher social coupling, we enter the $\bar{\text{D}}$C phase. Here, ostracism and defector payoff balance only when both are low because subsystem 1 is dominated by defectors. When it is instead dominated by cooperators, ostracism exceeds defector payoff due to the moderate social coupling. Finally, near the upper bound of the weak social coupling range, we reach the purely cooperative C phase. Here, ostracism exceeds defector payoff for all subsystem 1 population compositions, necessitating collapse of defection. The overall trend is that increasing the social coupling, even within its weak range, causes successive phase transitions in the subsystem 1 stability landscape, gradually increasing its cooperativity.

\end{document}